# Radiation Shielding Properties of $Nd_{0.6}Sr_{0.4}Mn_{1-y}Ni_yO_3$ Substitute with Different Concentrations of Nickle

**M. Kh. Hamad[a], M. H. A. Mhareb[b,c,*], Y. S. Alajerami[d,e], M. I. Sayyed[f], Gameel Saleh[g] Y. Maswadeh[h], Kh. A. Ziq[a]**

[a]*Physics Department, King Fahd University of Petroleum and Minerals, Dhahran 31261, Saudi Arabia*

[b]*Department of Physics, College of Science, Imam Abdulrahman Bin Faisal University, P.O. Box 1982, Dammam 31441, Saudi Arabia*

[c]*Basic and Applied Scientific Research Center, Imam Abdulrahman Bin Faisal University, P.O. Box 1982, Dammam 31441, Saudi Arabia*

[d]*Physics and Astronomy, Science Faculty, Ohio University, USA*

[e]*Medical Imaging Department, Applied Medical Sciences Faculty, Al Azhar University-Gaza*

[f]*Department of Physics, Faculty of Science, University of Tabuk, Tabuk, Saudi Arabia*

[g]*Department of Biomedical Engineering, College of Engineering, Imam Abdulrahman Bin Faisal University, P.O. Box 1982, Dammam, Saudi Arabia*

[h]*Department of Physics and Science of Advanced Materials Program, Central Michigan University, Mt. Pleasant, MI 48859, United States*

*****Corresponding author:** Dr. M. H. A. Mhareb (mhsabumhareb@iau.edu.sa)

### Abstract

In this work, we investigate the effect of Ni concentration on several shielding properties of $Nd_{0.6}Sr_{0.4}Mn_{1-y}Ni_yO_3$ ($0.00 \leq y \leq 0.20$) perovskite ceramic for possible use as radiation shielding materials. X-ray diffraction (XRD) analysis revealed that these ceramics have the orthorhombic structure with group space Pnma over a wide range of Ni-substitutions. Moreover, the analysis showed a nearly linear decrease in the lattice parameters and the unit cell volume (V) causing a gradual increase in the packing density with increasing Ni concentration. The shielding features for photons, neutrons, and protons of all ceramic samples were assessed. The mass attenuation coefficient (MAC) was computed at 0.1, 0.6, 1.25, 5 and 15 MeV by utilizing (MCNP) (version

5.0); the results were compared with the corresponding values obtained using Phy-X and XCOM program. The results obtained showed slight enhancement with increasing Ni contents. The substitution of Ni leads to progressive enhancement in effective removal cross-section of fast neutron ($\sum_R$) values. Whereas the values of Mass Stopping Power (MSP) and projected range for the protons showed a gradual reduction with increasing Ni concentration. These findings suggest that the current ceramic samples can be useful as radiation shielding materials.

## 1- Introduction

For example, reducing the exposure time, increasing the distance between radiation sources, and using proper and suitable shielding between the individuals and radiation source. High atomic number material (high density) such as Lead (Pb) is commonly used as a shield to protect from ionizing radiation such as gamma and X-ray. However, toxicity and health hazards for many materials (lead for example) necessitated the use of other materials such as glasses and ceramics for shielding applications (Gaikwad et al., 2018a; Bagheri et al., 2018; Al-Hadeethi and Sayyed, 2020; Junior et al., 2017). One of the key factors for choosing proper shielding materials is the gamma rays attenuation coefficients which largely depend on the density of these materials. The attenuation coefficient has been measured for several materials such as glass, concrete, granite, polymer and ceramic (Najam et al., 2016; Al-Hadeethi et al., 2020a; Al-Hadeethi et al., 2020b; Akkurt et al., 2005; El-Khayatt et.al., 2010; El-Khayatt and Akkurt 2013; Akkurt and Elkhayat A 2013; Kaky et al, 2020; Sayyed et al., 2020; Alya Abdsalam et al., 2020).

Ceramics have been extensively used in several medical and industrial applications as they are stable at high temperature, high resistance to oxidation, low thermal expansion, and low permittivity (Zhang et al., 1998). Ceramics are non-organic compounds with several attractive properties such as durability, hardness, toughness, modulus, lack of porosity, and environmentally friendly (non-toxic). Because of these properties, the ceramic composites are receiving increasing interests to be used in the radiation shielding field.

Amritphale et al., (2007a) used an aluminum industry waste (known as red mud or RMSM) as a shielding material. RMSM and barium compounds have been formed by phosphate bonding using the ceramic processing route. The X-ray attenuation characteristics of the proposed bauxite-red radiopaque material exhibited much less "half-value thickness (HVT)" compared to the conventional concrete and lead, for the various tube potentials (100 -250 KV) of X-ray photons (Amritphale et al., 2007a). Moreover, celsian ceramics prepared from fly ash, from the combustion of pulverized coal, have been developed as X-ray radiation shielding materials (Amritphale et al., 2007b). The fly ash was mixed with a barium compound to form a new fly ash radiopaque material (FARM) using the ceramic processing route and phosphate bonding. RMSM and FARM prepared ceramic samples showed the distinctive properties and these samples can be used properly for shielding in radiographic imaging rooms to protect against X-ray. New ceramic composites based on boron carbide ($B_4C$), were evaluated as a radiation shielding material for pulsed neutron scattering instrumentation (M. Celli et al., 2006). Other researchers have studied the gamma-ray buildup factor of ceramic/ceramic hosts such as magnesium diboride ($MgB_2$), titanium carbide (TiC), boron nitride (BN), silicon carbide (SiC), ferrite ($Fe_3O_4$). The exposure buildup factor for apatite, zirconolite and other ceramic hosts was investigated by T. Singh et al., 2013; V.P. Singh et al., 2018. The effective properties of shielding for another type of ceramic, mullite-barite ceramic, were studied by Ripin et al., 2018. Radiation tests showed that the best shielding ability of ceramics where the radiation attenuation between 99.11% and 97.42% was high at the tube potential 70 kV. Jawad et al., 2019 introduced different glazed and unglazed ceramics as shielding materials and measured their linear attenuation coefficient using two radiation sources: $^{60}$Co, and $^{137}$Cs. It was concluded that the glazed ceramics exhibited better attenuation of gamma radiation than the unglazed. Glazed ceramics may show reduced porosity and enhanced density making them more effective in radiation shielding. Akman et al. 2019 studied the shielding properties of various types of ceramics (silicide, boride and oxide ceramics) using transmission geometry at different energies values. The study showed that the lowest and highest values of the mean free path (MFP) and half-values layer (HVL) were discerned for titanium dioxide (TiO2) and magnesium silicide ($Mg_2Si$), respectively.

In this work, a series of the new perovskite ceramic $Nd_{0.6}Sr_{0.4}Mn_{1-y}Ni_yO_3$ were prepared in order to investigate the shielding properties. We also investigate the effects of Ni-substitution on various shielding properties. X-ray diffraction measurements have been used to obtain various physical properties. The calculated parameters are based on the mass attenuation coefficient extracted from Monte Carlo N-Particle Transport Code System (MCNP) (version 5.0), XCOM, and Phy-X program.

## 2- Material and methods

Perovskite ceramics of nominal composition $Nd_{0.6}Sr_{0.4}Mn_{1-y}Ni_yO_3$ with y = 0, 0.1, 0.15 and 0.2 and referred to as C1, C2, C3 and C4 respectively in this paper. The solid-state reaction technique has been used to prepare the stoichiometric ceramic samples (A.R. West. 2014). High purity (4N to 5N) oxides of Neodymium oxide ($Nd_2O_3$), strontium carbonate ($SrCO_3$), manganese oxide ($MnO_2$), and nickel oxide (NiO) were mixed and ground into powder in an agate mortar and pestle for about 30 minutes. The resulting powder was pressed into pellets and sintered at 1000 °C for 24 hours, and then at 1300 °C for another 10 hours (M. Kh. Hamad et al., 2018; M. Kh. Hamad et al. 2019). Table 1 shows the composition ratio for the ceramic samples. The phase purity and the lattice parameters were calculated from x-ray diffraction patterns. Bruker x-ray diffractometer with Cu Kα (λ=1.54056 Å) has been used to obtain the XRD-patterns at room temperature. The structures were examined between 20° to 80° using FULLPROF software (R.A. Young 1995). The electronic balance (Melter Toledo) with a sensitivity of $10^{-4}$ g was utilized to measure the density (ρ) based on Archimedes' principle using toluene as a buoyant liquid.

From the density data, the molar volume ($V_m$) can be computed using:

$$V_m = \frac{M}{\rho} \qquad (1)$$

Here, M is the average molecular weight for ceramic samples. On the other hand, there are some significant structural parameters such as packing density ($V_t$). This parameter was deduced from molar volume data and calculated using the following equations:

$$V_t = \sum_i \frac{V_i x_i}{V_m} \qquad (2)$$

Where $V_i$ and $x_i$ represent the packing factor, and the mole fraction of the oxides respectively. $V_i$ can be obtained by Pauling's ionic radii (L. Pauling 1940). While the Poisson's ratio (σ) values can be calculated from the $V_t$ data by utilizing the following equation:

$$\sigma = 0.5 - \left(\frac{1}{7.2V_t}\right) \qquad (3)$$

In this paper, Monte Carlo N-Particle Transport Code System (MCNP) (version 5.0) was conducted to estimate the MAC of the newly prepared ceramics. Previous studies have reported the efficiency of MCNP to evaluate radiation shielding parameters (M.G. Dong et al. 2017; H.O. Tekin and U. Kara 2016; K.A. Mahmoud 2019). The simulation was conducted in a sphere of 100 cm radius filled with dry air ($\rho = 1.205 \times 10^{-3}$ g cm$^{-1}$). A point source with a mono-energetic beam emission was assumed as the radiation source, and exposed perpendicular to the front surface of ceramic samples (in the Y-axis direction). To reduce possibility of background radiation, lead shielding collimators were assumed to be around the ionization chamber and ceramic sample as shown in Fig. 1. The irradiation process was conducted with and without ceramic samples to estimate the value of $I/I_o$. The simulation was performed based on the amount of Ni and repeated three times for each sample (C1-C4). The radiation source was proposed as a point source and the required commands such as energy (ERG), type of radiation (PAR), position (POS) and direction (DIR) were fully defined accordingly. The molecular weights of each constituent in the ceramic samples were obtained from the XCOM database program.

In the current simulation, the required data for simulation processes derived from the evaluated nuclear data file (ENDF), advanced computational technology initiative (ACTI), evaluated nuclear data library (ENDL), evaluated photon data library (EPDL), activation library (ACTL) and evaluations from the nuclear physics (T–16) group at Los Alamos. The energy detection in the NaI detector was determined by estimating the mesh tally (F4). The number of histories was increases (NPS > $10^7$), and geometry was simplified to reduce errors and variance of variance (VOV). The simulation was monitored by utilizing the CUT-OFF option to eliminate irradiation of low energy (< 1 keV).

The obtained results of MCNP-5 code were compared with the results of XCOM (Berger and Hubbel (1987) and Phy-X (Sakar, et al. 2020). In addition, several shielding properties such as MAC, mean free path (MFP), half-value layer (HVL), electron density, effective atomic number ($Z_{eff}$), energy absorption build-up factor (EABF), exposure build-up factor (EBF) have been investigated. MSP and projected range for the protons were evaluated of all ceramic samples. While $\sum_R$ was evaluated of all prepared ceramic samples. The shielding features were computed by utilizing several formulas listed in former work (Mhareb et al., 2019).

**3- RESULTS AND ANALYSIS**

Fig. 2 presents XRD patterns of all prepared ceramic samples. Refinement analysis of the crystalline structure shows that all samples are consistent with the standard orthorhombic pattern with Pnma (62) space group. However, there is a presence of a minor cubic perovskite phase in all samples. The cubic phase is commonly found in high-temperature treatment of these samples (M. Kh. Hamad et al. 2019). The peaks are shifted to higher values of 2θ with increasing the Ni concentration in the ceramic (inset of Fig. 2). This shift demonstrates a steady decrease in the lattice parameters (Table 2). The refined lattice parameters for all samples are shown in Table 2. The results show a nearly linear decrease in the parameters (a, b, and c), and the unit cell volume (V) with increasing Ni concentration. Moreover, this leads to a gradual increase in the packing density ($V_t$) of the ceramics with increasing Ni concentration.

Some physical and structural properties such as molar volume ($V_m$), packing density ($V_t$), and Poisson's ratio (σ) were calculated. These parameters are listed in Table 3. The packing density values show a significant increment with substitute the Ni at the Mn sites. This increase is in line with the gradual reduction in the molar volume values. This increment in ceramic density can be attributed to substitute light element (Mn) by heavy element (Ni). Table 3 also lists the Poisson's ratio which is a measure of the rigidity of the prepared ceramic samples. The Poisson's ratio varies between 0-0.5 for most materials. When the Poisson's ratio value was above 0.3, the materials can be classified by low cross-linking density. While the materials with high cross-linking density have Poisson's ratio values below 0.3 (V.V. Gowda et al., 2005). The values of Poisson's ratio for the ceramic samples are above 0.4, indicating low cross-linking for the current ceramic samples. It is observed slight variation in the Poisson's ratio values confirms the rigidity for ceramic samples with substitute the Ni instead of Mn.

The intensities of photons (both I and $I_0$) that pass through the ceramics have been enlisted in Table 4. These values helped us to simulate the MAC values for the investigated C1-C4 samples at 0.1, 0.6, 1.25, 5 and 15 MeV. In Table 5, we summarized the MAC values for the selected ceramics obtained by three methods namely MCNP5 code, Phy-X, and XCOM computer programs. Moreover, the MAC (both MCNP-5 and XCOM findings) for the present ceramics is plotted in three-dimensional figure as a function of the photon energy and the Ni-concentration as shown in Fig. 3. This is important to validate the simulated input file and to test the accuracy in the simulated μ/ρ. The data presented in Table 5 indicate a qualitative agreement within acceptable variance between various results obtained by MCNP5 code and data obtained by Phy-

 For example, for C1 and at 0.1 MeV, the MAC value obtained by the simulation method is 1.0191 cm$^2$/g, while the corresponding value obtained by Phy-X software is 0.9312 cm$^2$/g and that evaluated by XCOM is 0.9447 cm$^2$/g.

In order to understand the effect of the energy and the content of Ni on the radiation attenuation features for the prepared ceramics, we plotted the XCOM results in the energy range of 0.015 to 15 MeV in Fig. 4. This figure shows that the MAC values for C1, C2, C3 and C4 ceramics decrease with the increasing the energy. This is mainly due to several photon interaction processes existing at different energy zones (M.I. Sayyed et al., 2020). The results in Fig. 4 showed a significant reduction in MAC values at low energies. For C1 (as an example), the MAC changes from about 50 to 1.5 cm$^2$/g between these energies. This is due to the photoelectric effect which is dominant at low energy. The chances of occurrence of this mode of interaction change with the energy as $E^{-3.5}$. Also, this mode of interaction changes with the atomic number as $Z^{4-5}$, therefore the expected maximum values of MAC occurred in this zone. The maximum MAC occurs at 0.015 MeV and equals to 49.890, 50.460, 50.747 and 51.034 cm$^2$/g for C1, C2, C3 and C4 respectively. In the second energy zone namely between 0.15 and 1 MeV, the Compton scattering is energetically possible. Due to this process, MAC changes slightly between 0.570 and 0.070 MeV (for C1). Also, due to this process, we can see that the composition of the ceramics is not affecting the μ/ρ values. For example, at 0.2 MeV, the MAC values for C1, C2, C3 and C4 respectively are 0.3137, 0.3148, 0.3153 and 0.3158 cm$^2$/g respectively. For the photon energy larger than 1 MeV, the pair production becomes very important and accordingly the MAC becomes almost constant with energy. This is in agreement with the recent findings reported by different groups (M.I. Sayyed et al., 2019; M. Kurudirek et al., 2018).

The HVL for the C1, C2, C3 and C4 ceramics is plotted in Fig. 5. Evidently, for all tested ceramics, the HVL increases as the energy increases. The increase is dramatically steep for energies bellow about 1 MeV followed by gradual increases, reaching maximum value near 10 MeV. The minimum HVL is reported at 15 MeV and equals to 0.0027, 0.0025, 0.0023 and 0.0022 cm for C1, C2, C3 and C4 respectively. The discontinues in the HVL at 0.05 MeV is attributed to the K-absorption edge of Nd (Z=60). At 0.1 MeV, the HVL values for the tested ceramics are 0.0884, 0.0846, 0.0777 and 0.0726 cm. While, at 1 MeV the HVL increases to 2.2022, 2.1148, 1.9474 and 1.8214 cm for C1-C4. This implies that it is important to increase the thickness of the ceramics in case of utilizing the ceramics in applications required high photon energies. Also, Fig. 5 shows that the HVL of these ceramics changes in order as C1>C2>C3>C4. This is related to the amount of Ni in the ceramics. The replacement of Mn by Ni causes an increase in the density (see Table 3), therefore the HVL decreases with the addition of Ni. The HVL curves reveal that an improvement in the radiation attenuation features for the tested ceramics with the addition of Ni.

The MFP for the C1-C4 ceramics is plotted as a function of the energy in Fig. 6-A, while in Fig. 6-B we plotted the MFP as a function of the density at some selected energies namely 0.1, 0.6, 1.25, 5 and 15 MeV. The minimum MFP for the four tested ceramics was noticed for E<0.1MeV. It lies within the range of 0.0031-0.0039 cm at 15 keV and 0.1047-0.1276 cm at 0.1 MeV. For E>0.1 MeV, the MFP increases rapidly as can be noticed from Fig. 6-A. This means that the MFP has the same HVL trend and it is in line with the results reported for some materials such as glasses (Ozge Kilicoglu and H.O. Tekin, 2020). Beyond 1 MeV, the MFP shows an energy-dependent behavior. This suggests that as the energy increases, the probability of radiation interaction with the C1, C2, C3, and C4 ceramics decreases, and hence more photons can penetrate the tested ceramics. From Fig. 6-B, increasing the density of the tested ceramics leads to a decrease in the MFP. This result suggests that increasing the Ni content in the ceramics (which affects the density) can reduce the thickness required to shield the radiation to a specific value. For example, the density changes from 5.182 g/cm$^3$ (for C1) to 6.258 g/cm$^3$ (for C4) and the MFP values of these two samples are 2.294 and 1.896 cm (this is at 0.6 MeV). At 5 MeV, the MFP of both ceramics is 5.707 cm (for C1) and 4.707 cm (for C4). These results are in line with the results for other ceramics such as silicide, boride and oxide ceramics (Akman et al, 2019).

The gamma radiation shielding feature for the tested ceramics is investigated in term of the effective atomic number (Z$_{eff}$). In Fig. 7 we plotted the Z$_{eff}$ for the ceramics under study as a function of photon energy and the content of Ni. This figure shows that the Z$_{eff}$ for all test ceramics decreases with increasing the energy (except at 0.05 MeV). Several theoretical and experimental works have reported the same evidences for the dependency of Z$_{eff}$ upon the energy like rocks (Obaid, Shamsan S., et al., 2018), and glasses (D. K. Gaikwad, et al., 2018). Also, from Fig. 7 we can conclude that the lowest and highest values of $Z_{eff}$ are corresponding to C1 and C4 respectively. As C4 contains the maximum amount of Ni (y=0.2), this explains the highest $Z_{eff}$ for this ceramic. This emphasizes our findings in the previous curves that the addition of Ni improves the attenuation ability for the tested ceramics and C4 can effectively absorb more photons than C1, C2 and C3. The $Z_{eff}$ values at 0.015 MeV varied between 45.90 and 46.01, at 0.05 MeV varied between 53.86 and 53.88 (and these are the maximum $Z_{eff}$ values reported for these ceramics), while at 1 MeV the $Z_{eff}$ takes the following values: 20.23, 20.42, 20.51 and 20.61 for C1, C2, C3 and C4 respectively. For these ceramics, the $Z_{eff}$ reaches the minimum values at 1.5 MeV and equals to 19.99, 20.81, 20.27 and 20.37. For E>1.5 MeV we can see that the $Z_{eff}$ increases with increases the energy due to the domination of pair production. For instance, the $Z_{eff}$ for C1 changes from 20.24 to 29.31 between 2 and 15 MeV and from 20.62 to 29.77 for C4. Additionally, Fig. 8 presents the relation between the $Z_{eff}$ and the effective electron density. It is evident that the $Z_{eff}$ changes linearly with the effective electron density which means that both quantities have similar behavior with the energy.

The variation of EBF and EABF with the energy for C4 (as an example) at some penetration depths has been plotted in Fig. 9 (a for EBF and b for EABF). C1, C2 and C3 have similar shape given in this figure. Obviously, the value of both parameters for all penetration depths increases with the increase of the energy up to 0.05 MeV and sudden jumps occur at this energy and these can be explained by the k edge absorption of Nd as we mentioned in the HVL curves. Thereafter, the EBF and EABF increase and reached the maximum at about 1 MeV, then again decrease in slight rate up to around 8 MeV and then both parameters increase quickly especially at 40 mfp. The present trend in these parameters can be demonstrated according to three important mode of radiation interaction with the matter as discussed in detail by Manjunatha and Rudraswamy (H.C. Manjunatha et al., 2012). It is important to mention that at for E>8 MeV and for 30 and 40 mfp, both EBF and EBAF have relatively high values. For example at 10 MeV, the EBF is 290.19 (at 40 mfp), while the EABF is 201.59. Increasing the penetration depth of the tested ceramics causing an increase of the thickness which in turn results in increasing the scattering events in the ceramic samples and this is the reason for the large EBF and EABF values.

In addition, the variation of $\sum_R$ with the content of Ni is represented graphically in Fig. 10. The results show that C1 and C4 have the lowest and largest values of $\sum_R$ respectively. The addition of Ni leads to enhancement in the neutron shielding ability since the $\sum_R$ increases with the addition of Ni content. The $\sum_R$ values for the tested ceramics are 0.109, 0.113, 0.122 and 0.130 cm$^{-1}$. Recently, Kaçal et al. (Mustafa R. Kacal, et al., 2018) prepared different ceramics and calculated $\sum_R$ for these ceramics and they reported the following values for $\sum_R$: peridot (0.0983 cm$^{-1}$), aluminum nitride (0.1152 cm$^{-1}$), yttrium oxide (0.0881 cm$^{-1}$), ruby (0.1248 cm$^{-1}$), silicon nitride (0.1225 cm$^{-1}$) and magnesium silicate (0.1145 cm$^{-1}$). From these results, we can see that our tested ceramics have better neutron shieling properties than peridot and yttrium oxide, while C4 has better neutron shielding ability than ruby silicon nitride and magnesium silicate.

The Mass Stopping Power (MSP) elucidates the rate of energy loss from incident particle through the medium. This parameter is very important to show the amount of energy created in a specific area per 1 gram in the prepared ceramics. In the current study, we used the SRIM code with the ESTAR database offered by the National Institute of Standard and Technology (NIST) (Ziegler et al., 2008; SRIM.org). Fig. 11 exhibits the calculated MSP of the prepared ceramics at a different range of proton (H) energies (0.01-10 MeV). In all prepared ceramics, the MSPs increased with increased increase kinetic energy of the proton. In addition, it is clear that the MSPs are in an opposite relation with the concentrations of Ni, where C4 (0.2 of Ni) has the lowest MSP.

Fig. 12 shows the calculated projected range of the prepared ceramics in case of proton irradiation. This parameter is significant to determine the effective shielding material by express how far the proton can penetrate and at what depth can rest. This thickness will increase with increasing energy of the incident proton. The current ceramics exhibit very promising thickness based on the tested kinetic energy (0.01-10 MeV), particularly for C4.

### 4- Conclusion

By using a conventional solid-state reaction, four samples of $Nd_{0.6}Sr_{0.4}Mn_{1-y}Ni_yO_3$ (0.00≤y≤0.20) perovskite were fabricated to evaluate the shielding properties for photons, neutrons, and protons. The XRD results affirm the crystallinity nature and the major phase is orthorhombic structure for current ceramic samples. The values of packing density, MAC and $\sum_R$ exhibit significant increment with addition Ni to the ceramic system. Moreover, the addition of Ni inversely affected the MSP and the projected range for the protons. The obtained results showed good shielding properties for photons, neutrons, and protons. This improvement in shielding features indicates the ability to use prepared ceramics samples as good shielding materials.

## Acknowledgment

We acknowledge the support provided by Deanship of scientific research, King Fahd University of Petroleum & Minerals, Kingdom of Saudi Arabia.

Table 1. The chemical formula and density for the prepared ceramic samples.

| Sample ID | Chemical formula | Density (g/cm$^3$) |
|---|---|---|
| C1 | $Nd_{0.6}Sr_{0.4}MnO_3$ | 5.182 |
| C2 | $Nd_{0.6}Sr_{0.4}Mn_{0.90}Ni_{0.10}O_3$ | 5.393 |
| C3 | $Nd_{0.6}Sr_{0.4}Mn_{0.85}Ni_{0.15}O_3$ | 5.855 |
| C4 | $Nd_{0.6}Sr_{0.4}Mn_{0.80}Ni_{0.20}O_3$ | 6.258 |

Table 2. Lattice parameters of all prepared ceramic samples

| | Ceramic Sample's Code | | | |
|---|---|---|---|---|
| | C1 | C2 | C3 | C4 |
| $a(\text{Å})$ | 5.47(0) | 5.46(5) | 5.45(8) | 5.45(3) |
| $b(\text{Å})$ | 7.67(5) | 7.66(0) | 7.65(2) | 7.64(7) |
| $c(\text{Å})$ | 5.43(6) | 5.42(8) | 5.42(0) | 5.41(4) |
| $V(\text{Å}^3)$ | 228.19 | 227.23 | 226.34 | 225.74 |

Table 3. Physical and structural properties of all prepared ceramic samples

| Measurements | Ceramic Sample's Code | | | |
|---|---|---|---|---|
| | C1 | C2 | C3 | C4 |
| $V_m$ (cm$^3$ mol$^{-1}$) | 33.565 | 32.139 | 29.550 | 27.598 |
| Packing density ($V_t$) | 0.416 | 0.435 | 0.473 | 0.507 |
| Poisson's ratio ($\sigma$) | 0.444 | 0.443 | 0.442 | 0.442 |

Table 4. Linear and mass attenuation coefficient results at specific energies (MCNP5)

|  | Energy (MeV) | I (count) | FSD | Io (count) | FSD | Linear Att. (cm$^{-1}$) | Mass Att. (cm$^2$/g) |
|---|---|---|---|---|---|---|---|
| **C1** | 0.1 | 125719 | 0.022 | 866385 | 0.033 | 4.8257 | 0.9447 |
|  | 0.6 | 743991 | 0.013 | 876709 | 0.034 | 0.4103 | 0.0799 |
|  | 1.25 | 823799 | 0.044 | 916385 | 0.017 | 0.2662 | 0.0524 |
|  | 5 | 879710 | 0.045 | 937179 | 0.028 | 0.1561 | 0.0301 |
|  | 15 | 935333 | 0.033 | 995977 | 0.045 | 0.1571 | 0.0303 |
| **C2** | 0.1 | 114298 | 0.054 | 866385 | 0.036 | 5.0638 | 0.9389 |
|  | 0.6 | 738057 | 0.050 | 876709 | 0.038 | 0.4304 | 0.0787 |
|  | 1.25 | 818797 | 0.048 | 916385 | 0.019 | 0.2815 | 0.0513 |
|  | 5 | 874219 | 0.023 | 937179 | 0.024 | 0.1717 | 0.0308 |
|  | 15 | 931366 | 0.012 | 995977 | 0.045 | 0.1677 | 0.0311 |
| **C3** | 0.1 | 95716 | 0.035 | 866385 | 0.036 | 5.5073 | 0.9212 |
|  | 0.6 | 725804 | 0.027 | 876709 | 0.036 | 0.4722 | 0.0806 |
|  | 1.25 | 810104 | 0.045 | 916385 | 0.018 | 0.3081 | 0.0517 |
|  | 5 | 871072 | 0.048 | 937179 | 0.029 | 0.1808 | 0.0308 |
|  | 15 | 924289 | 0.051 | 995977 | 0.045 | 0.1867 | 0.0319 |
| **C4** | 0.1 | 82011 | 0.050 | 866385 | 0.036 | 5.8936 | 0.9312 |
|  | 0.6 | 714601 | 0.037 | 876709 | 0.038 | 0.5111 | 0.0787 |
|  | 1.25 | 801803 | 0.019 | 916385 | 0.019 | 0.3339 | 0.0523 |
|  | 5 | 866055 | 0.022 | 937179 | 0.025 | 0.1952 | 0.0312 |
|  | 15 | 918302 | 0.025 | 995977 | 0.048 | 0.2029 | 0.0314 |

FSD (Fractional Standard Deviation)

**Table 5** Mass attenuation coefficients results at specific energies (Phy-X, MCNP and XCOM)

| Energy (MeV) | | C1 | C2 | C3 | C4 |
|---|---|---|---|---|---|
| 0.1 | Phy-X | 1.0191 | 1.0211 | 1.0230 | 1.0261 |
| | MCNP5 | 0.9312 | 0.9389 | 0.9402 | 0.9412 |
| | XCOM | 0.9447 | 0.9503 | 0.9531 | 0.9559 |
| | Δ | 1.43 | 1.21 | 1.36 | 1.55 |
| 0.6 | Phy-X | 0.0842 | 0.0855 | 0.0866 | 0.0901 |
| | MCNP5 | 0.0791 | 0.0799 | 0.0806 | 0.0817 |
| | XCOM | 0.0808 | 0.0809 | 0.0810 | 0.0811 |
| | Δ | 2.13 | 1.24 | 0.49 | 0.74 |
| 1.25 | Phy-X | 0.0551 | 0.0570 | 0.0580 | 0.0585 |
| | MCNP5 | 0.0514 | 0.0521 | 0.0527 | 0.0533 |
| | XCOM | 0.0534 | 0.0536 | 0.0537 | 0.0538 |
| | Δ | 3.82 | 2.84 | 1.88 | 0.93 |
| 5.0 | Phy-X | 0.0341 | 0.0346 | 0.0361 | 0.0368 |
| | MCNP5 | 0.0302 | 0.0308 | 0.0308 | 0.0312 |
| | XCOM | 0.0318 | 0.0319 | 0.0320 | 0.0321 |
| | Δ | 5.16 | 3.51 | 3.82 | 2.84 |
| 15 | Phy-X | 0.0350 | 0.0372 | 0.0382 | 0.0389 |
| | MCNP5 | 0.0303 | 0.0311 | 0.0319 | 0.0324 |
| | XCOM | 0.0318 | 0.0320 | 0.0321 | 0.0323 |
| | Δ | 4.83 | 2.85 | 0.62 | 0.31 |

Δ=[(XCOM-MCNP5)/((XCOM+MCNP5)/2)]*100%

**Figure captions**

**Figure 1.** Total simulation setup for irradiation the new ceramic samples.

**Figure 2.** X-ray diffraction patterns for prepared ceramic samples at room temperature. The peak is shifted to the right by increasing the Ni concentrations (inset).

**Figure 3.** Comparison of XCOM and MCNP5 mass attenuation coefficients for the prepared ceramics.

**Figure 4.** Three-dimension image for mass attenuation coefficient (XCOM).

**Figure 5.** The Half Value Layer (HVL) for the investigated ceramics.

**Figure 6.** The Mean Free Path (MFP) of prepared ceramics: A) Dependence of MFP on incident photon energy; B) The effect of ceramic density on the MFP at different energies.

**Figure 7.** Variation of effective atomic number ($Z_{eff}$) as a function of photon energy for the prepared ceramics.

**Figure 8.** Variation of $Z_{eff}$ with $N_{eff}$ for the prepared ceramics at mutli-energetic photons.

**Figure 9.** Variation of EBF and EABF values as a function of photon energy of C4 sample.

**Figure 10.** The removal cross-section for the prepared ceramics samples

**Figure 11.** Variation of mass stopping powers (MSP) for proton interaction for the prepared ceramics.

**Figure 12.** Variation of the projected range with kinetic energy for photon interaction of the prepared ceramics.

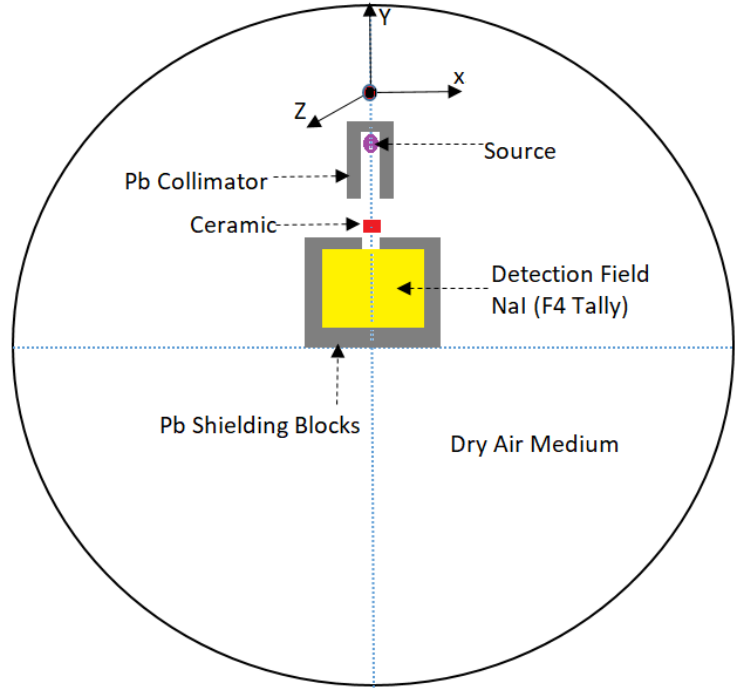

**Figure 1.**

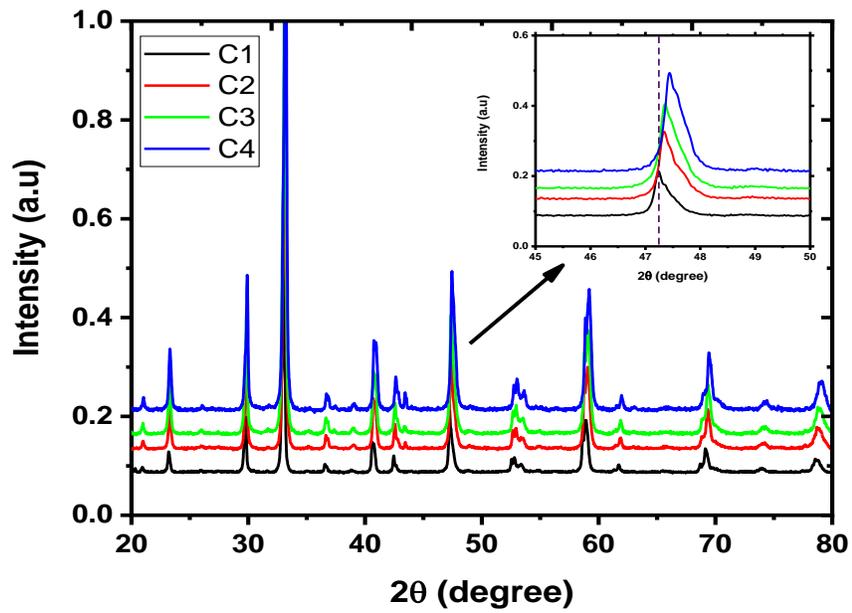

**Figure 2.**

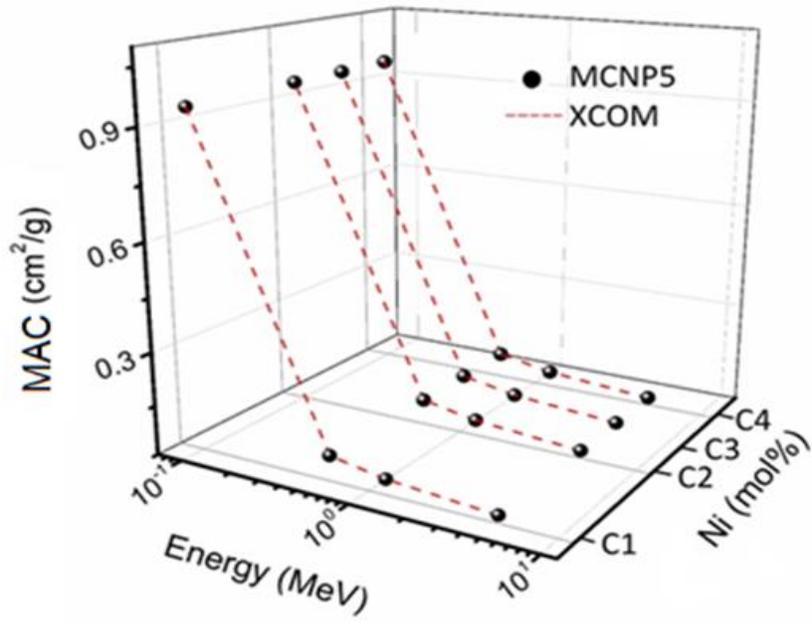

**Figure 3.**

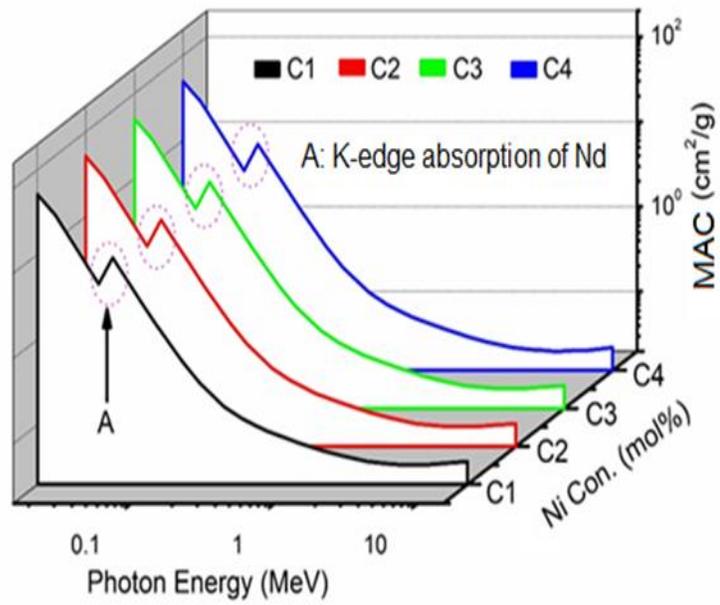

**Figure 4.**

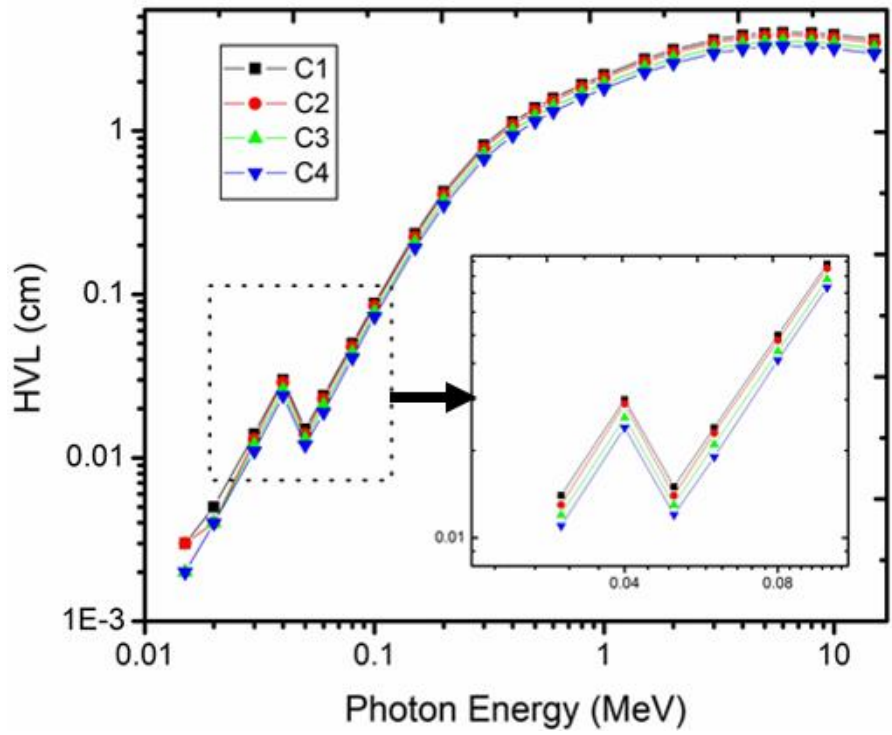

**Figure 5.**

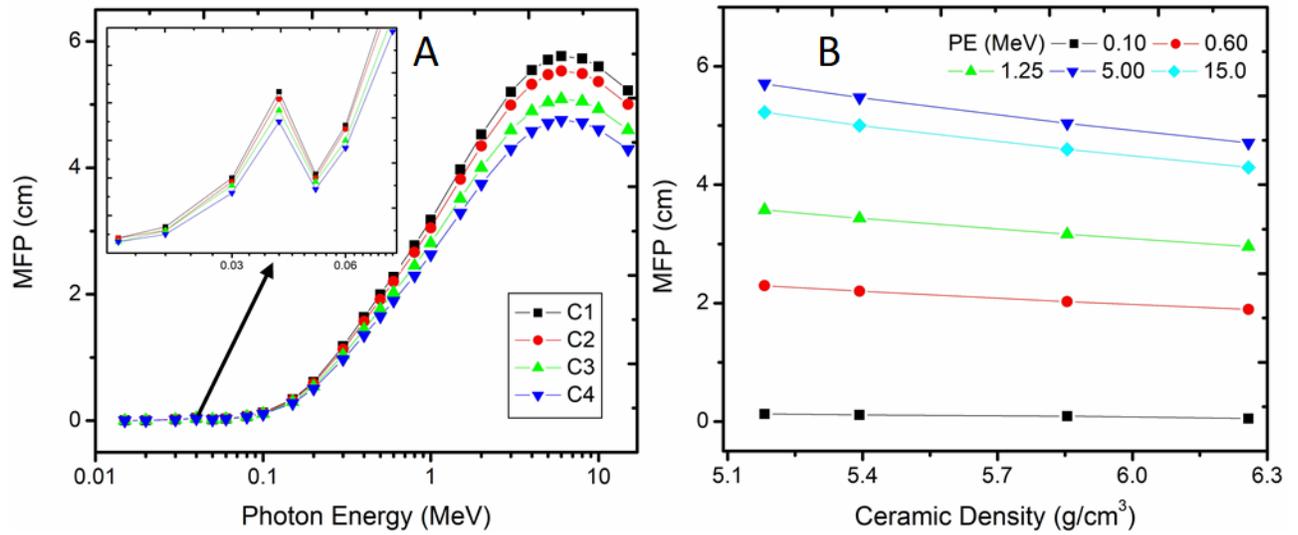

**Figure 6.**

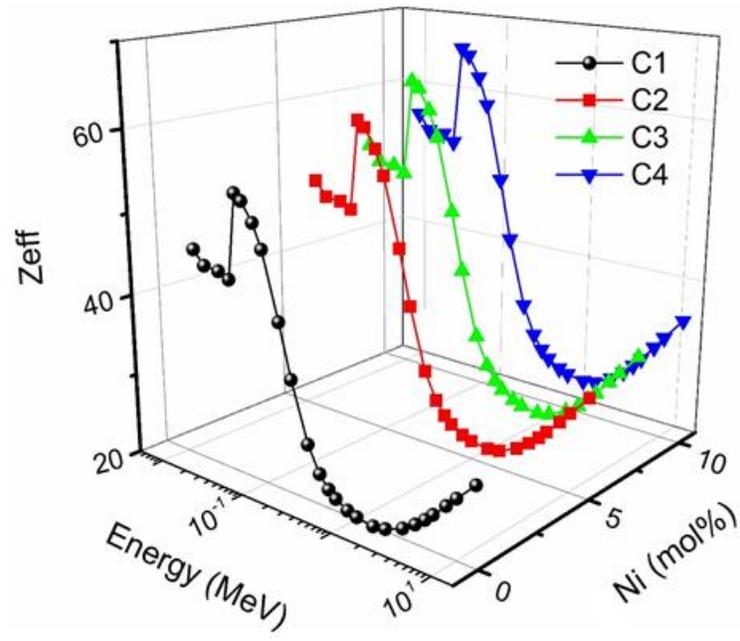

**Figure 7.**

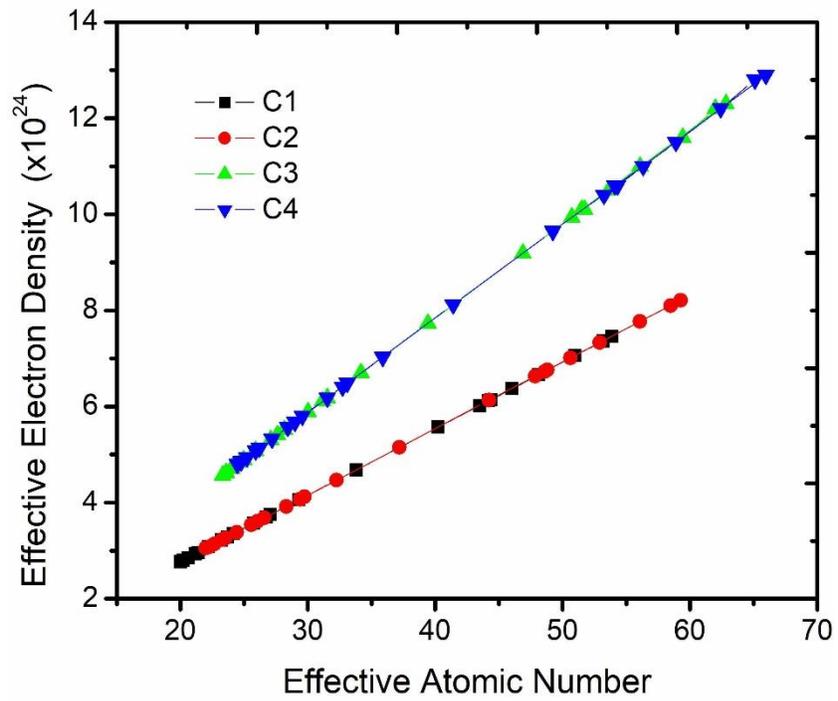

**Figure 8.**

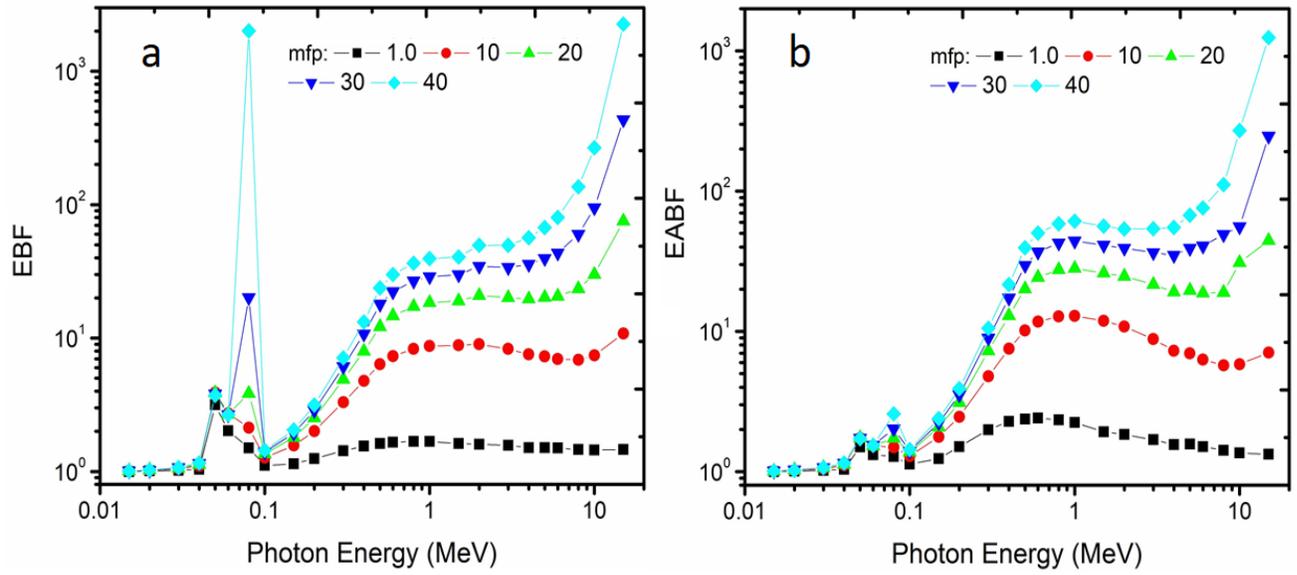

**Figure 9.**

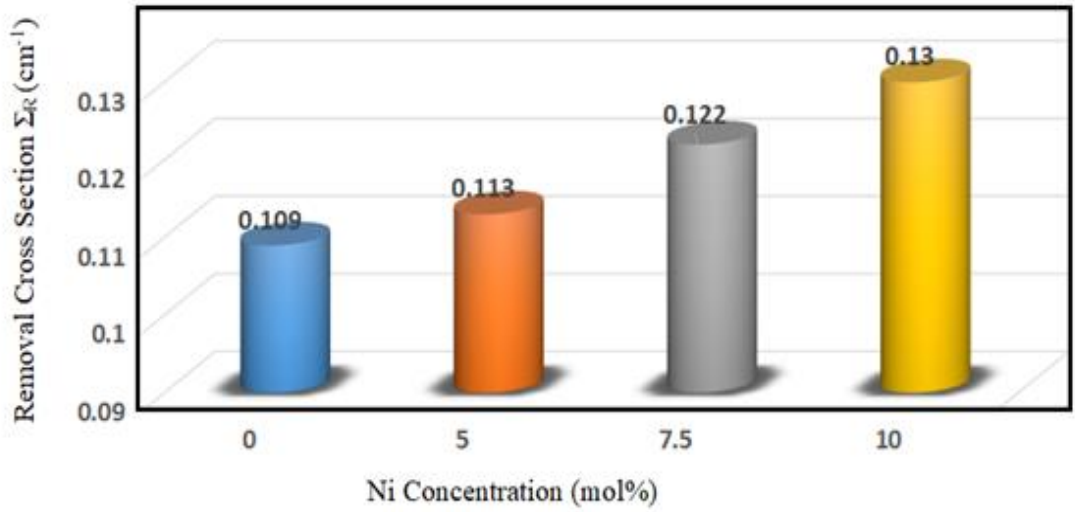

**Figure 10.**

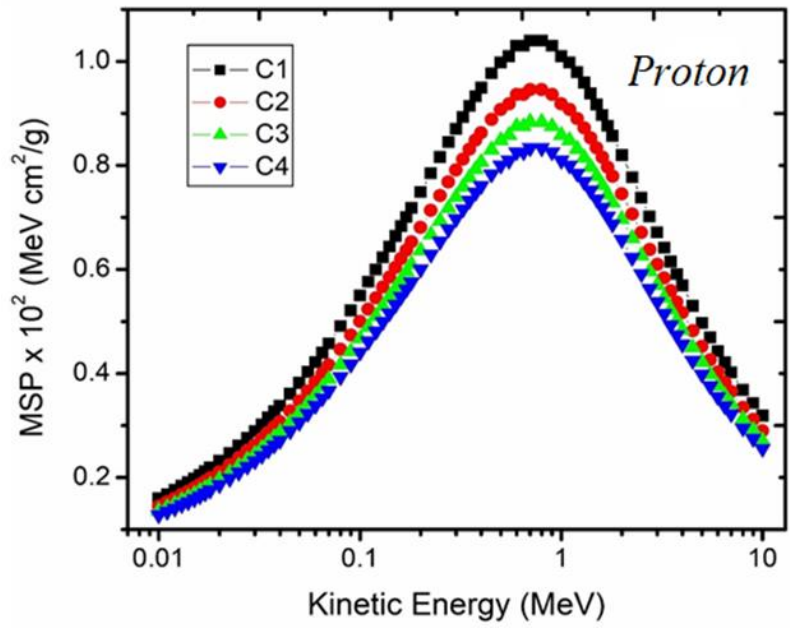

**Figure 11.**

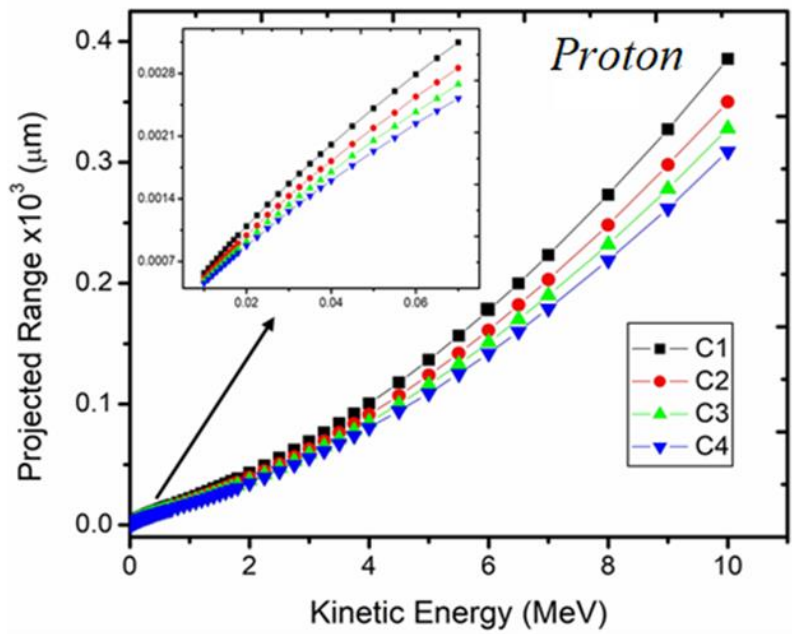

**Figure 12.**